\documentclass[a4paper,preprint]{revtex4}
\usepackage{times}
\usepackage{graphicx}
\usepackage{url}
\usepackage{longtable}
\usepackage{amsmath}
\usepackage{color}

\def \gsim{\mathrel{\mathpalette\@versim>}}
\def \lsim{\mathrel{\mathpalette\@versim<}}
\def \neut{\tilde \chi^0}
\def \charg{\tilde \chi^+}
\def \glue {\tilde g}
\def \st {\tilde t}
\def \sq {\tilde q}
\def \sb {\tilde b}

\begin{document}
\title{Constraints on supersymmetry with light third
family from LHC data}
\author{Nishita Desai} 
\affiliation{Harish-Chandra Research Institute,
  Chhatnag Road, Jhunsi, Allahabad - 211 019, India} 
\author{Biswarup Mukhopadhyaya}
\affiliation{Harish-Chandra Research Institute, Chhatnag Road, Jhunsi, Allahabad -
  211 019, India}
  \preprint{HRI-RECAPP-11-009}

\begin{abstract}
We present a re-interpretation of the recent ATLAS limits on
supersymmetry in channels with jets (with and without b-tags) and
missing energy, in the context of light third family squarks, while
the first two squark families are inaccessible at the 7 TeV run of the
Large Hadron Collider (LHC). In contrast to interpretations in terms
of the high-scale based constrained minimal supersymmetric standard
model (CMSSM), we primarily use the low-scale parametrisation of the
phenomenological MSSM (pMSSM), and translate the limits in terms of
physical masses of the third family squarks. Side by side, we also
investigate the limits in terms of high-scale scalar non-universality,
both with and without low-mass sleptons. Our conclusion is that the
limits based on zero-lepton channels are not altered by the mass-scale
of sleptons, and can be considered more or less model-independent.
\end{abstract}
\maketitle

\section{Introduction}
The Large Hadron Collider (LHC) at CERN is now running and the two
experiments ATLAS and CMS are hard at work in constraining theories
beyond standard model (BSM).  Supersymmetry (SUSY)
\cite{Nilles:1983ge, Martin:1997ns} has long been one of the most
popular BSM models, due to its ability to solve the hierarchy problem
and to provide a dark matter candidate in its R-parity conserving
versions.  The minimal supersymmetric standard model (MSSM) extends
the nineteen standard model parameters to over a hundred.  However,
the proposed mechanisms of SUSY breaking, together with the
requirements of suppresing flavour-changing neutral currents (FCNC)
and obtaining the correct electroweak breaking scale often suggest a
common origin of some of the parameters at high scale.  This results
in highly constrained models like the minimal supergravity model
(mSUGRA) \cite{Chamseddine:1982jx, Arnowitt:1992aq} or the very
similar constrained MSSM (cMSSM) \cite{Kane:1993td}, gauge mediated
SUSY \cite{Giudice:1998bp} and others.  Alternately, the new
parameters can be constrained by demanding only that all observed
experimental constraints are obeyed, thus arranging SUSY breaking mass
parameters in such a way as to suppress flavour-changing neutral
current (FCNC) processes. This leads to a phenomenological MSSM
(pMSSM) \cite{Berger:2008cq} which has nineteen parameters -- the
gaugino mass parameters $M_1$, $M_2$ and $M_3$; the squark mass
parameters for the first two generations $M_Q$, $M_U$ and $M_D$, the
third generation squarks $M_{3Q}$, $M_{BR}$, $M_{TR}$; the
corresponding slepton mass parameters $M_L$, $M_R$, $M_{\tau L}$,
$M_{\tau R}$; the trilinear couplings $A_t$, $A_b$ and $A_\tau$; and
the higgs sector parameters $\mu$, $M_A$ and $\tan \beta$, the ratio
of the vacuum expectation values of the two Higgs doublets.

The best reach in superparticle masses at the LHC is expected in the
channel with two or more hard jets and missing energy
\cite{Aad:2009wy} which is the characteristic signature from $\glue
\glue$ and $\sq \glue$ production.  In particular, the simplest decays
of the gluino and the squark, viz.\ $\glue \rightarrow qq \tilde
\chi^0_1$ and $\sq \rightarrow q \tilde \chi^0_1$ result in four and
three-jet states with large missing energy for the two production
processes respectively.  Also, the sub-dominant $\sq \sq^*$ and $\sq
\sq$ production processes would result in two-jet final states.
Therefore, looking for 2-4 jets with missing energy is known to be the
best channel for SUSY searches.  However, in the case that the first
two generations of squarks are not accessible at the LHC (as can
indeed be motivated from suppression of flavour changing neutral
currents)\cite{Brax:2000ip,Dimopoulos:1995zw}, the power of these
searches would be dramatically reduced.  Most importantly, the
t-channel $\sq \glue$-type processes which contribute largest in the
standard cMSSM analysis, would be severely suppressed because of the
miniscule fraction of b and t-quarks in the proton.  Similarly, other
t-channel processes leading to $\sq \sq$, $\sq \sq^*$ are also
suppressed.  The dominant production processes are therefore $\glue
\glue$ and s-channel $\sq \sq^*$ ($\sq = \st_{1,2},\sb_{1,2}$).

Since the limits on the mass of third generation squarks in a
cMSSM-based analysis follow simply from limits obtained from squark
and gluino production processes, they cannot be considered truly
indicative of the limit on stop and sbottom masses.  We therefore
reinterpret the ATLAS limits in the pMSSM model where the first two
generations of squarks and all sleptons are decoupled.

Considerable interest has also grown in recent times in SUGRA with
non-universal high-scale masses.  A high-scale parametrisation has the
advantage that the masses of several particles are obtained naturally
through renormalisation group (RG) running.  Here too, we focus of
situations where the third family sfermions are within the reach of
the LHC, while the first two families are heavy. This, among other
things, helps in a natural suppression of FCNC.  The advantage of this
scenario will be to allow us to investigate the effect of a low-mass
slepton sector without requiring its full pMSSM parametrisation.

We base our study on the data from the ATLAS experiment for signatures
with jets and missing energy with and without b-tagged jets
\cite{daCosta:2011qk, Aad:2011ks, ATLAS-CONF-2011-086, Aad:2011ib,
  ATLAS-CONF-2011-098}.  The ATLAS analysis assumes an mSUGRA-type
unification for the interpretation of its data.  As mentioned above,
the limits from this analysis cannot be applied to the third
generation squarks in a model independent way.  The results have been
interpreted in terms of a high-scale non-universal model in
\cite{Sakurai:2011pt}.  However, its dependence on mSUGRA based mass
hierarchies (e.g. the lighter stop $\st_1$ is always right-handed, the
lightest neutralino is mostly bino-like etc.)  hampers full
understanding of the implication of the experimental data on the third
generation squark sector.  We therefore perform a more detailed study
by performing a low-scale pMSSM analysis with a scan over the physical
stop/sbottom masses.  We also include the case of stop decay via the
flavour-violating decay $\tilde t \rightarrow c \tilde \chi^0_1$
\cite{Hikasa:1987db, Boehm:1999tr, Muhlleitner:2011ww} when all other
decays are forbidden by kinematics.

The phenomenology of third generation squarks has also been studied in
various scenarios \cite{Kadala:2008uy, Graesser:2008qi,
  Bhattacharyya:2008tw, Desai:2009ex, Beenakker:2010nq, Baer:2010ny,
  Bartl:2010du, Li:2010zv,Bornhauser:2010mw, Huitu:2011cp,
  Datta:2011ef,Kats:2011qh,Essig:2011qg,Bi:2011ha}.  It should be
noted that while this manuscript was being prepared, similar studies
on related issues \cite{Papucci:2011wy, Brust:2011tb} have also
appeared in literature.

The paper is structured as follows: in the next section, we describe
our simulation strategy and demonstrate the degree of agreement with
the original ATLAS analysis.  In section \ref{sec:phen}, we describe
the phenomenological modelling of the stop and sbottom sectors and
present the results corresponding to this scenario.  The case of
various high-scale non-universal scenarios is described in section
\ref{sec:high-scale}.  We present
our conclusions in section \ref{sec:conclusions}.

\section{CMSSM: Simulation of the signal and ATLAS exclusion curves}
\label{sec:simulation}

We simulate the signal using {\sc Pythia} 6.4 \cite{Sjostrand:2006za}
and all strong production cross sections are normalised to their
next-to-leading order (NLO) values as obtained from Prospino 2.1
\cite{Beenakker:1996ed}.  The renormalisation and factorisation scales
are set to the average of masses in the final state of the hard
scattering process.  We follow the detector acceptance region for all
reconstructed objects and apply all the cuts as described in the ATLAS
papers.

The full set of identification and acceptance cuts is as follows:

\begin{itemize}
\item {\bf Electrons:} (1) $p_T> 20$ GeV (2) $|\eta|<2.47$ (3) Sum of
  $p_T$ of particles within $\Delta R = \sqrt{(\Delta \eta)^2+(\Delta
  \phi)^2} < 0.2$ should be less than 10 GeV (4) Event vetoed if
  electron found in $1.37<|\eta|< 1.52$.
\item {\bf Muons:} (1) $p_T> 20$ GeV (2) $|\eta|<2.4$ (3) Sum of $p_T$
  of charged tracks within $\Delta R = \sqrt{(\Delta \eta)^2+(\Delta
    \phi)^2} < 0.2$ should be less than 1.8 GeV.
\item {\bf Jets:} (1) Formed using Anti-kt algorithm from Fastjet with
  parameter $R=0.4$ (2) $p_T > 20$ GeV.
\item {\bf b-tagged jets:} A jet is b-tagged if a b-hadron falls
  within a cone of radius $R$ from a jet.  We have checked that this
  reproduced the 50 \% tagging efficiency for $t \bar t$ samples as
  mentioned in ~\cite{ATLAS-CONF-2011-098}.
\item Missing transverse energy is calculated by summing over the
  $p_T$ of all objects and all stable visible particles not belonging
  to any reconstructed objects but falling within $|\eta|<4.9$ with
  $p_T> 0.5$ GeV.
\end{itemize}

To account for detector effects, we smear the momenta of leptons and
jets obtained from the Monte Carlo generator according to
\begin{equation}
\frac{\sigma(E)}{E} =  \frac{a}{\sqrt E} \oplus b
\end{equation}
The values of $a$ and $b$ are (0.11, 0.007) for electrons, (0.03,
0.06) for muons and (1.0, 0.05) for jets \footnote{B.\ Mellado,
  private communication.}.  After smearing, we apply cuts used in each
ATLAS analysis under consideration.

\begin{figure*}
\includegraphics[width=80mm]{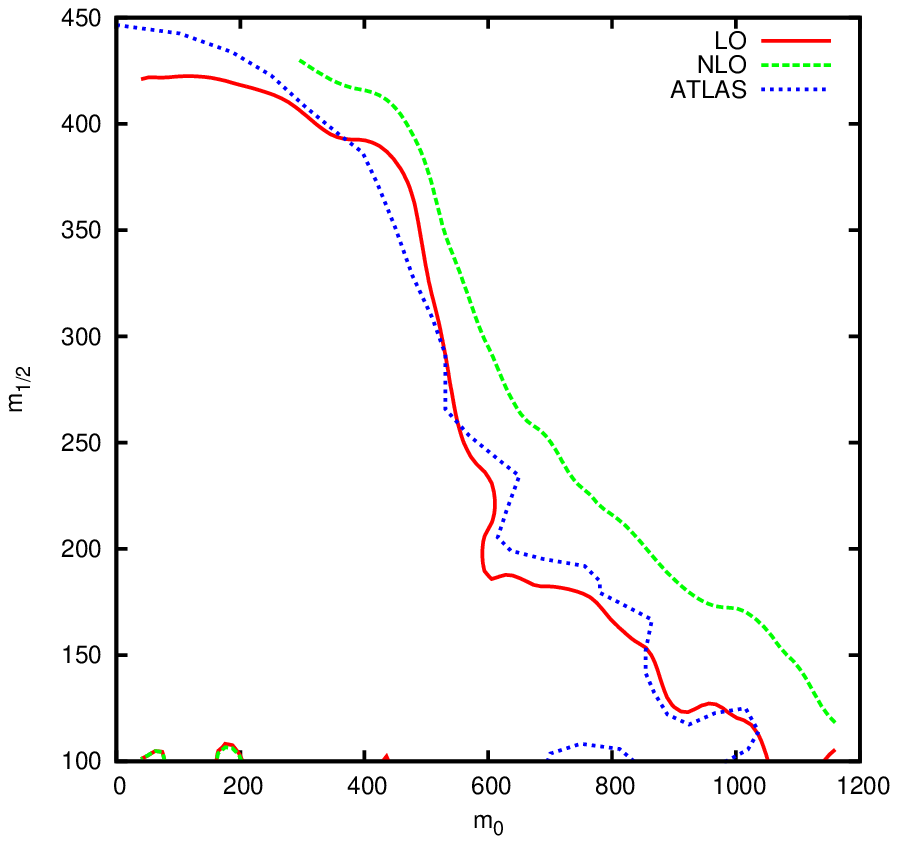}
\includegraphics[width=80mm]{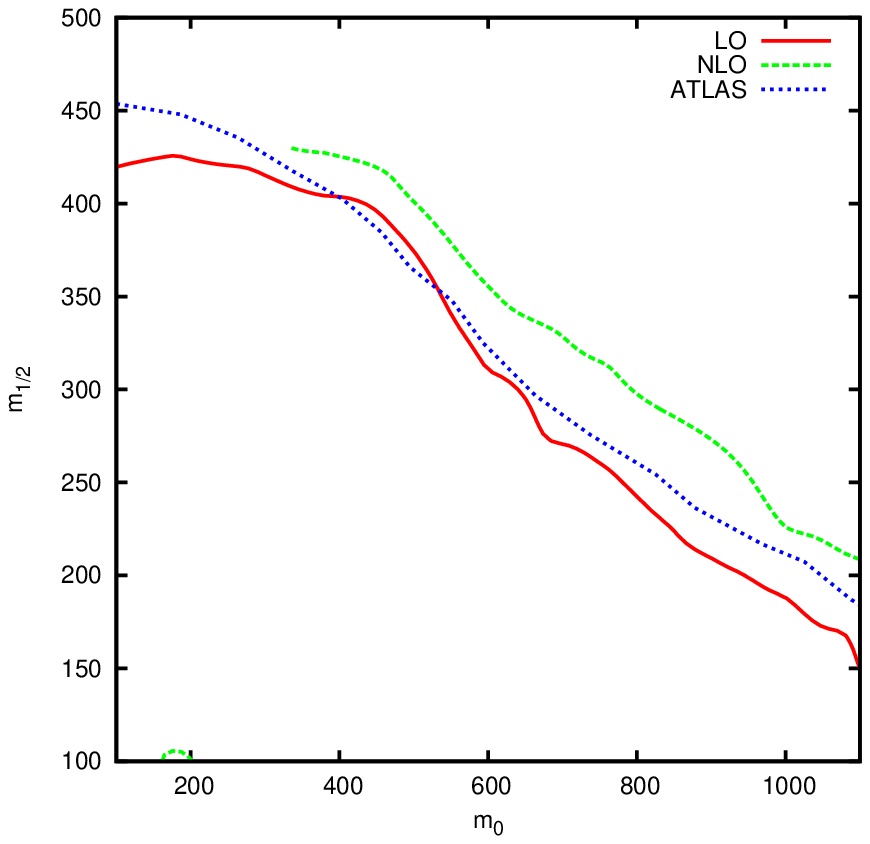}
\includegraphics[width=80mm]{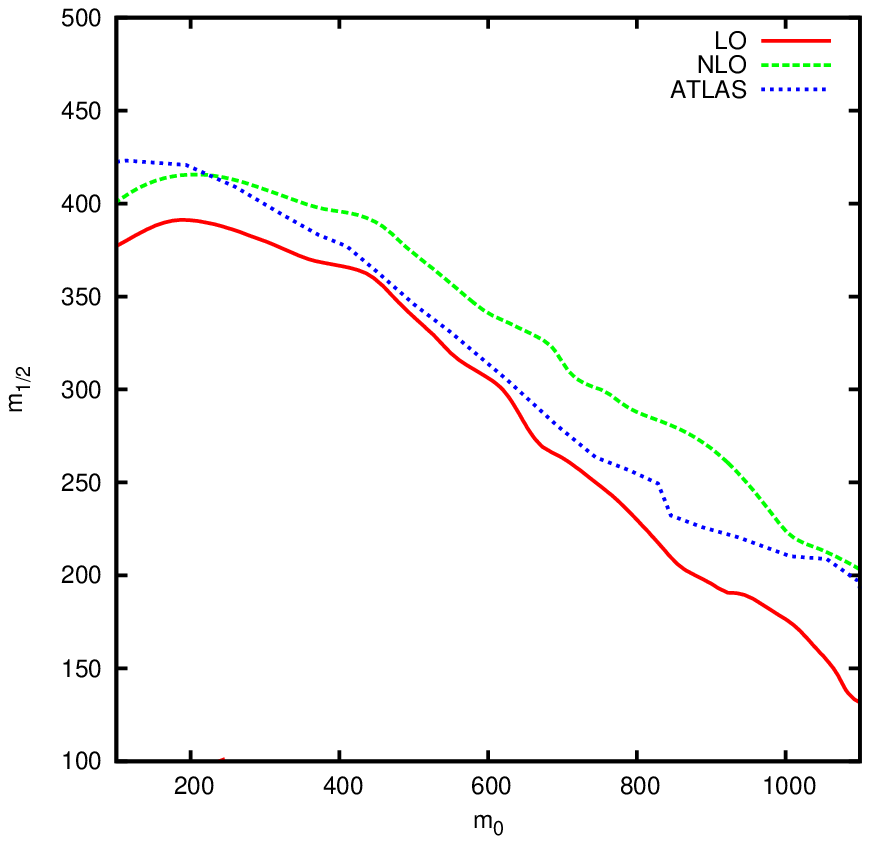}
\includegraphics[width=80mm]{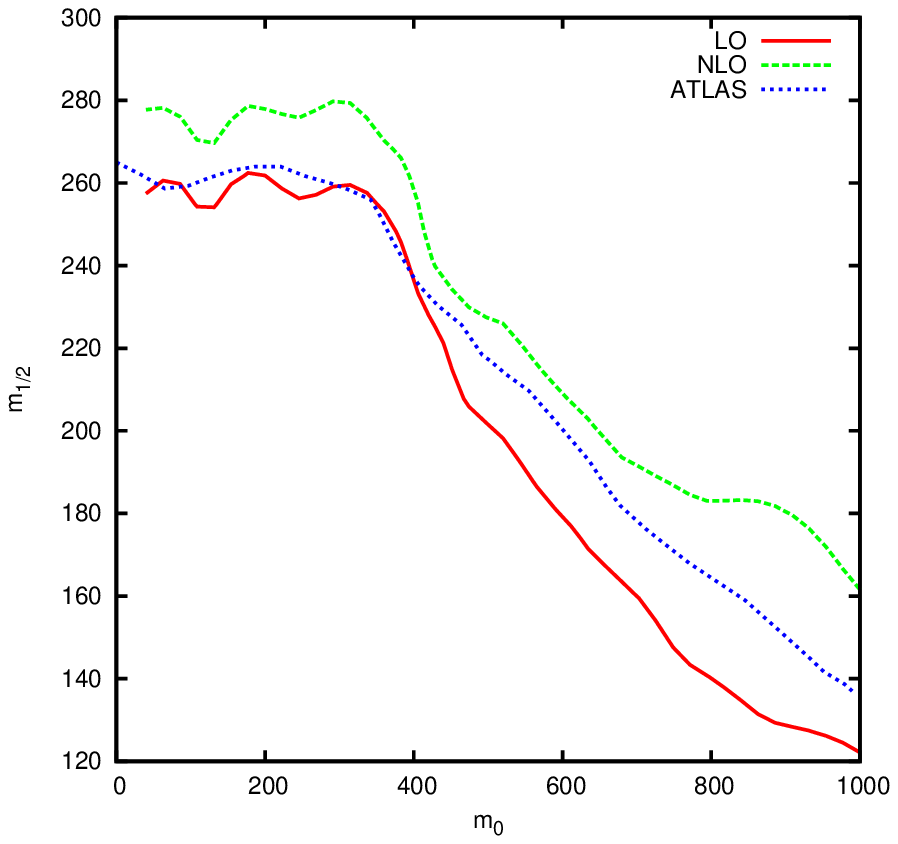}
\caption{\label{fig:atlas} Comparison of our exclusion curves with the
  ATLAS exclusion curves~\cite{ATLAS-CONF-2011-086, Aad:2011ks}.  The
  panels represent: 2-jet with 165 pb$^{-1}$ (top-left), 3-jet with
  165 pb$^{-1}$ (top-right), 4-jet with 165 pb$^{-1}$ (bottom-left)
  and b-jet with 0-leptons with 35 pb$^{-1}$ (bottom-right).  The true
  ATLAS curve lies between our LO and NLO contours in all cases.}
\end{figure*}

A cross-check of our analysis is the correct reproduction of the
missing transverse energy (MET) and effective mass
($M_{\mathrm{eff}}$) distributions and consequently the reproduction
of the ATLAS exclusion curves \cite{ATLAS-CONF-2011-086, Aad:2011ks}
in the context of mSUGRA.  The online resources for the jets+MET
analysis for 35 pb$^{-1}$ \cite{daCosta:2011qk} provide both the
benchmark points used in the scan as well as the efficiencies and
cross sections at each of the points.  We use this information to
verify the correctness of our simulation.  We present in
Fig.\ \ref{fig:atlas}, the final exclusion curves for the jets+MET
analysis at 165 pb$^{-1}$ which are obtained using the same acceptance
and smearing parameters.  We find that the ``true'' ATLAS curve lies
in between our leading order (LO) and NLO curves in all cases.

Since we aim to determine the limits on third generation squarks, the
limits on b-tagged events are of particular importance and are
expected to provide much stronger limits.  Therefore, reliable
modelling of b-tagging is of prime importance in looking at these
signals.  The work reporting the analysis of b-jet+MET describes the
performance of the b-tagging algorithm as having an efficiency factor
of 50\% for a $t \bar t$ sample.  We reproduce this number by this
simple algorithm: we first form jets using the anti-$k_T$ algorithm
using {\sc Fastjet} 2.4 \cite{Cacciari:2008gp} with the radius
parameter $R=0.4$.  A jet is assumed to be b-tagged if a b-hadron is
found within a distance $R$ from its axis.  Since the correct
reproduction of MET and $M_{\mathrm{eff}}$ has been verified from the
non-b-tagged samples, we can see that this algorithm gives a
reasonably correct b-jet tagging by looking at the bottom-right panel
in Fig.\ \ref{fig:atlas}.  Here too, we find that the LO and NLO
curves encapsulate the ATLAS curve reliably.

In all the above cases, we find that the LO curve only slightly
underestimates the ATLAS limits.  In the worst case, the difference
between the LO limits and the ATLAS curve is within 20\%.  We
therefore mostly present the LO mass limits in the subsequent
study. Our LO results with simplistic detector simulation do not
differ by more than 20\% from what a full re-analysis of the data,
including detector responses, would give. Our cross-checks on the
mSUGRA results convince us such limits are adequate for putting across
our main point, given the uncertainty of detector effects.

We use the results on jets+MET for 1.04 fb$^{-1}$~\cite{Aad:2011ib}
and b-jets+MET results for 0.83 fb$^{-1}$~\cite{ATLAS-CONF-2011-098}
for our analysis.  As mentioned above, our simulation includes all the
cuts in each of the channels under consideration.  The benchmark
points for our analyses are obtained using SUSPect 2.41
\cite{Djouadi:2002ze}.  To obtain the exclusion curves, we use the
value of cross section times acceptance provided by the ATLAS
analysis.  The values corresponding to different channels taken from
these ATLAS analyses are summarised in Table~\ref{tab:crossx}.  The
names of the signal regions are the same as those used in the
respective ATLAS analyses.

We concentrate here only on signals without leptons in the final
state.  The primary reason for this is that we wish to investigate the
third generation squark sector in as model independent a way as
possible.  Leptonic states generally result from decays on gauginos
into gauge bosons or into sleptons which then decay into leptons.  For
a pMSSM study based on leptonic signatures, it would therefore be
imperative to also include a completely general gaugino sector as well
as a low-mass slepton sector.  Since adding a completely
phenomenological slepton sector means adding five new parameters which
complicate the analysis beyond too much, our low-scale analysis deals
with a decoupled sleptonic sector.  It is, in principle, possible that
the limits obtained in the decoupled slepton limit are diluted when
decays into sleptons (and hence leptons) are possible.  We include 
the possibility  of a low-mass slepton sector  in section
\ref{sec:high-scale}, where we use high-scale parametrisation and allow
the RG running to determine the masses in the slepton sector.
However, we shall see that allowing low-mass sleptons do not make
significant difference to the limits for signatures based on jets and
missing energy.  

We perform our analysis retaining the cMSSM-like gaugino mass pattern
$M_1:M_2:M_3=1:2:6$.  In general, for most gaugino mass patterns, we
expect our results to remain fairly unchanged since our signal does
not depend strongly on particles obtained from intermediate decays in
SUSY cascades.  However, we explicitly comment at the end of the paper
on the extreme cases of gaugino mass patterns that would be likely
yield results drastically different from ours.
 
It is also possible to ask what fraction of the pMSSM phase space is
ruled out by current data.  For the effect of the experimental limits
on the full pMSSM space, we refer the reader to \cite{Conley:2011nn,
  Sekmen:2011cz}.  The effects on the cMSSM parameter space are
addressed in, for example,~\cite{Allanach:2011ut, Allanach:2011wi,
  Buchmueller:2011sw, Sekmen:2011cz} whereas other interpretations of
the recent LHC data on SUSY have been discussed in
\cite{Feldman:2011me, Athron:2011wu,Scopel:2011qt, Akula:2011zq}.

 \begin{table}[htdp]
\begin{center}
\begin{tabular}{lc}
\hline
\textbf{Channel} & $\sigma \times acc$ (fb) \\
\hline
2 jets + MET &  24 \\
3 jets + MET & 30 \\
4 jets + MET ($M_{\mathrm{eff}} = $1 TeV) & 32 \\
\hline
1 btag + $M_{\mathrm{eff}} > 500$ (3JA)  & 288 \\
1 btag + $M_{\mathrm{eff}} > 700$ (3JB)  & 61\\
2 btag + $M_{\mathrm{eff}} > 500$ (3JC)  & 78\\
2 btag + $M_{\mathrm{eff}} > 700$ (3JD)  & 17 \\
\hline
\end{tabular}
\end{center}
\caption{\label{tab:crossx} The values of cross section times
  acceptance from ATLAS analysis used for applying exclusion limits.
  The first set uses 1.04 fb$^{-1}$ of data~\cite{Aad:2011ib} whereas
  the second uses 0.833 fb$^{-1}$ of data~\cite{ATLAS-CONF-2011-098}.}
\end{table}

\section{Parameterisation of the stop and sbottom sector}
\label{sec:phen}
We work in the pMSSM framework where the parameters are assigned at
the low-scale.  The program {\sc SuSpect} is used to ensure that
electroweak symmetry breaking has correctly taken place and the
spectrum is consistent.  To start with, we retain the cMSSM-like
gaugino mass ratios, correspond to $M_1:M_2:M_3 \simeq 1:2:6$ among
the U(1), SU(2) and SU(3) gaugino masses.  The squark masses of the
first two generations and all the slepton masses are set to 2 TeV
which is beyond the reach of the 7 TeV LHC run.  The stop and sbottom
sector can each be described by three parameters -- the two mass
eigenstates and the mixing angle.  We use these as the input
parameters for the scan and use the diagonalisation to determine the
left and right handed mass parameters of the pMSSM. The stop sector
requires three parameters --- the masses $M_{\st_1}$ and $M_{\st_2}$
and the stop mixing angle $\theta_{\st}$.  Using the stop mass-squared
matrix diagonalisation condition
\begin{equation}
\left(
\begin{array}{cc}
 M^2_{\st_1} & 0  \\
  0 &  M^2_{\st_2}  \\   
\end{array}
\right) = \mathcal{R} 
\left(
\begin{array}{cc}
 M^2_{3Q} & m_t X_t \\
 m_t X_t  &  M^2_{TR}  \\   
\end{array}
\right) \mathcal{R}^{-1}~;~\mathcal{R} = \left(
\begin{array}{cc}
 \cos \theta_{\st} & \sin \theta_{\st}  \\
 -\sin \theta_{\st} &  \cos \theta_{\st}  \\   
\end{array}
\right)
\end{equation}
where $X_t = A_t - \mu \cot \beta$, we can use the low-scale masses
$M_{\st_1}$, $M_{\st_2}$ and $\theta_{\st}$ as the input parameters
which uniquely determine $A_t$ given $\mu$ and $\tan \beta$.  The left
handed sbottom mass is expected to be close to the left handed stop
mass since they are derived from the same parameter ($M_{3Q}$) in the
low-scale pMSSM model.  The right handed sbottom mass and $A_b$ can
then be set depending on the requirement of $M_{\sb_1}$, $M_{\sb_2}$
and $\theta_{\sb}$.

Assuming that the third generation squarks and the gluino are the only
strongly charged superparticles accessible at the LHC, we investigate
in particular, the following cases:
\begin{itemize}
\item Case A: $sin \theta_{\st} = 0.99$ i.e.\ $\tilde t_1 \simeq
  \tilde t_R$ is the lightest squark.  This is commonly the case in
  cMSSM models.  We also set $M_{\tilde b_1} \simeq M_{\tilde t_2}
  \simeq M_{\tilde b_2} = M_{\st_1} + 500$ GeV, which makes the
  sbottom sector somewhat heavier than the lighter stop.  We scan the
  parameter space in $M = M_{\st_1}$.

\item Case B: $sin \theta_{\st} = 0.01$ i.e.\ $\tilde t_1 \simeq t_L
  $, $\simeq \tilde b_1 \simeq \tilde b_L$.  The lightest stop and
  sbottom are nearly degenerate and mostly left handed.  $M_{\tilde
    t_2} \simeq M_{\tilde b_2} = M_{\st_1} + 500$ GeV.  We scan the
  parameter space in $M = M_{\st_1} = M_{\sb_1}$.

\item Case C: $\sb_1 \simeq \sb_R$ is the lightest squark.  $M_{\tilde
  b_2} \simeq M_{\tilde t_1} \simeq M_{\tilde t_2} = M_{\sb_1} + 500$
  GeV with $\sin \theta_{\st} = 0.70$ and $\sin \theta_{\sb} =
  0.99$. We scan the parameter space in $M = M_{\sb_1}$.

\item Case D: $M_{\tilde t_1} \simeq M_{\tilde b_1} \simeq M_{\tilde
  t_2} \simeq M_{\tilde b_2}$.  This is the case of maximal mixing in
  both stop and sbottom sectors.  The most stringent limits on the
  light third family scenario will arise for this particular case, as
  it allows all four squarks to be produced with similar cross
  sections.  The scan in this case is over the common mass of all the
  third generation squarks.
\end{itemize}

\begin{table}[htdp]
\begin{center}
\begin{tabular}{lc}
\hline
\textbf{Parameter} & \textbf{Scan range} \\
\hline
$M_{\st_1}$ & 100 - 2000 GeV \\
$M_2$       & 150 - 600 GeV\\
$\tan \beta$ & 5, 10, 40 \\
$\mu$ & -200, 200, 500, 1000 GeV\\
\hline
\end{tabular}
\end{center}
\caption{\label{tab:param} The parameters of the scan in $M_{\st_1}$-$M_2$ space.}
\end{table}%

In each case, we are now able to perform a scan over the $M-M_2$
plane, where $M$ is the mass of $\st_1$ in cases A, B and D and
$\sb_1$ in the case C.  The Tevatron reach for searches in the $\st_1
\rightarrow c \neut_1$ rule out stop masses up to 
180 GeV \cite{CDF-note-9834}.  We therefore start our search at
$M_{\st_1} =$ 200 GeV and scan up to 2 TeV.  Also, since we assume
gaugino mass unification, we use the chargino mass limits from
Tevatron ($M_{\charg_1} > $164 GeV) to start our scan at $M_2 =$ 150
GeV and vary $M_2$ up to a value of 600 GeV which would correspond to
gluino mass of 1.8 TeV and therefore cover the entire range of masses
reachable at the 7 TeV run of the LHC.  We fix $M_A = 400$ GeV and
perform this scan for 12 combinations of 3 values of $\tan \beta$ and
4 values of $\mu$, which are listed in Table
\ref{tab:param}.

\begin{figure*}[tb]
\begin{tabular}{cc}
\includegraphics[scale=.6]{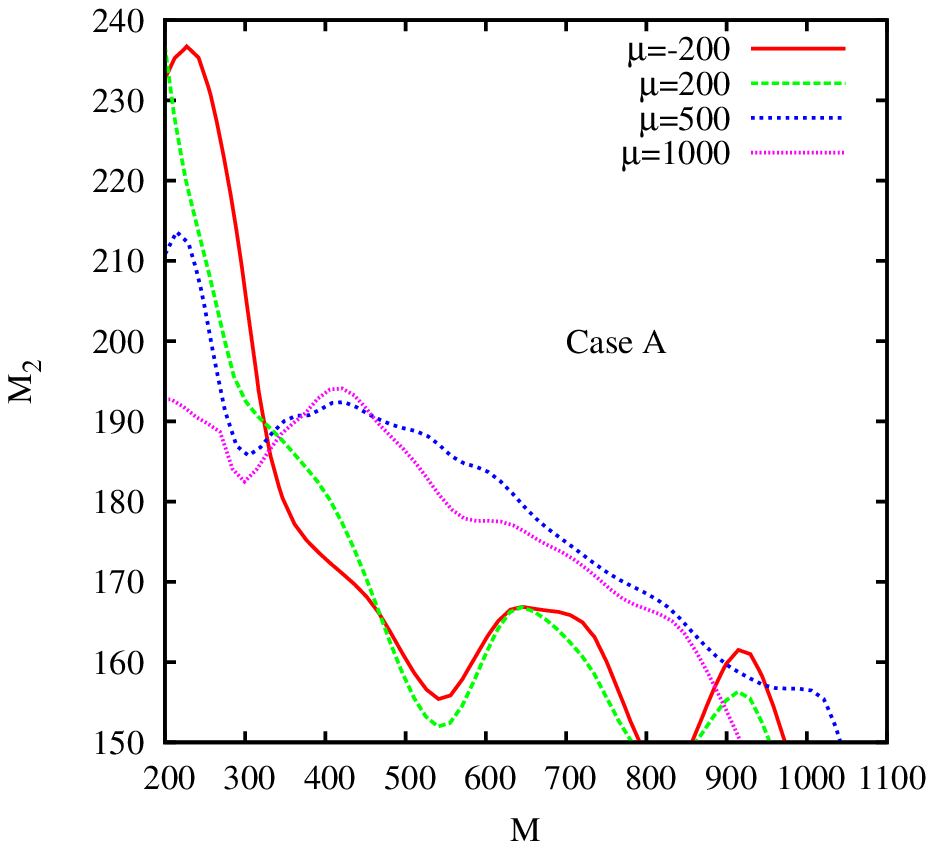} & 
\includegraphics[scale=.6]{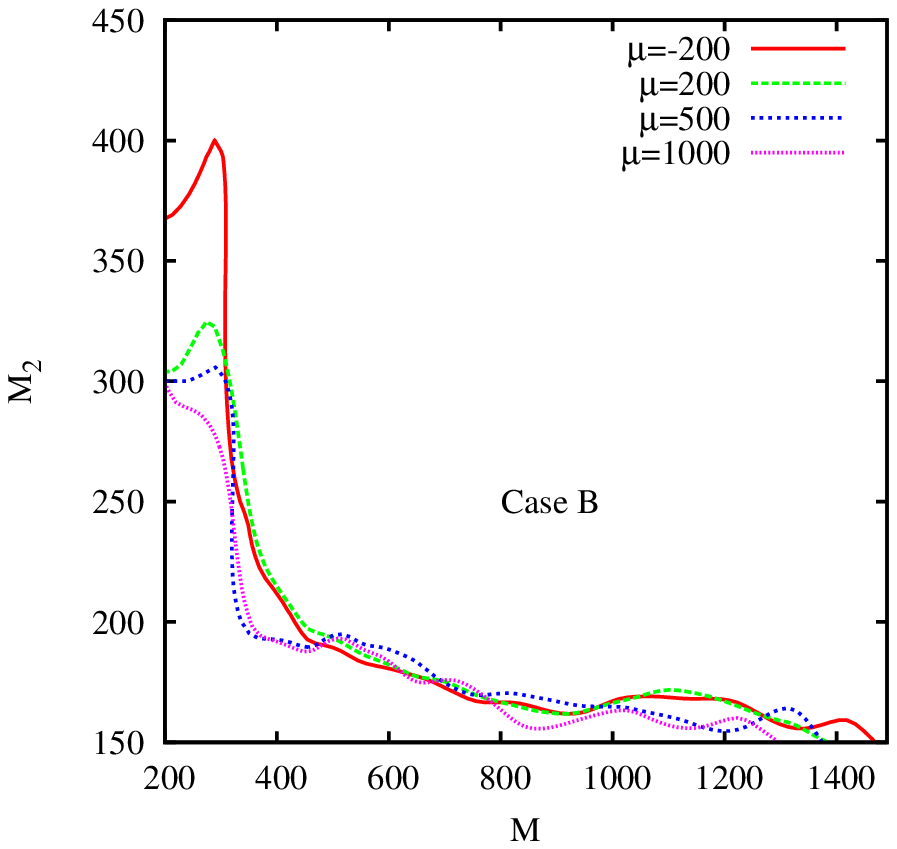} \\
\includegraphics[scale=.6]{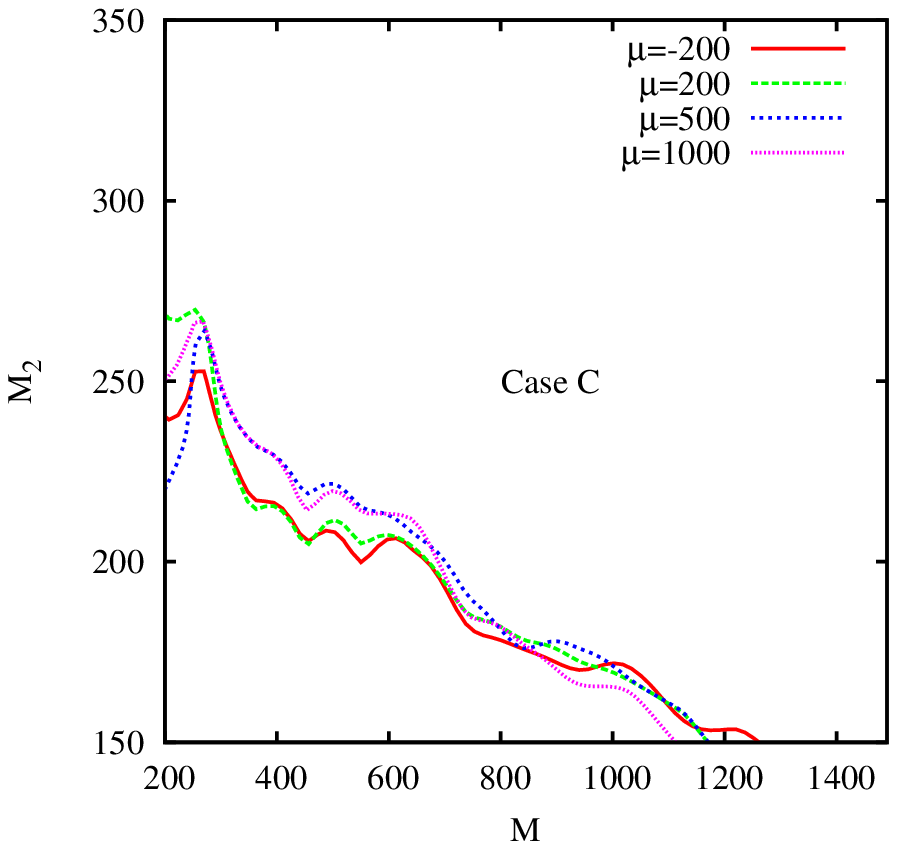} &  
\includegraphics[scale=.6, clip]{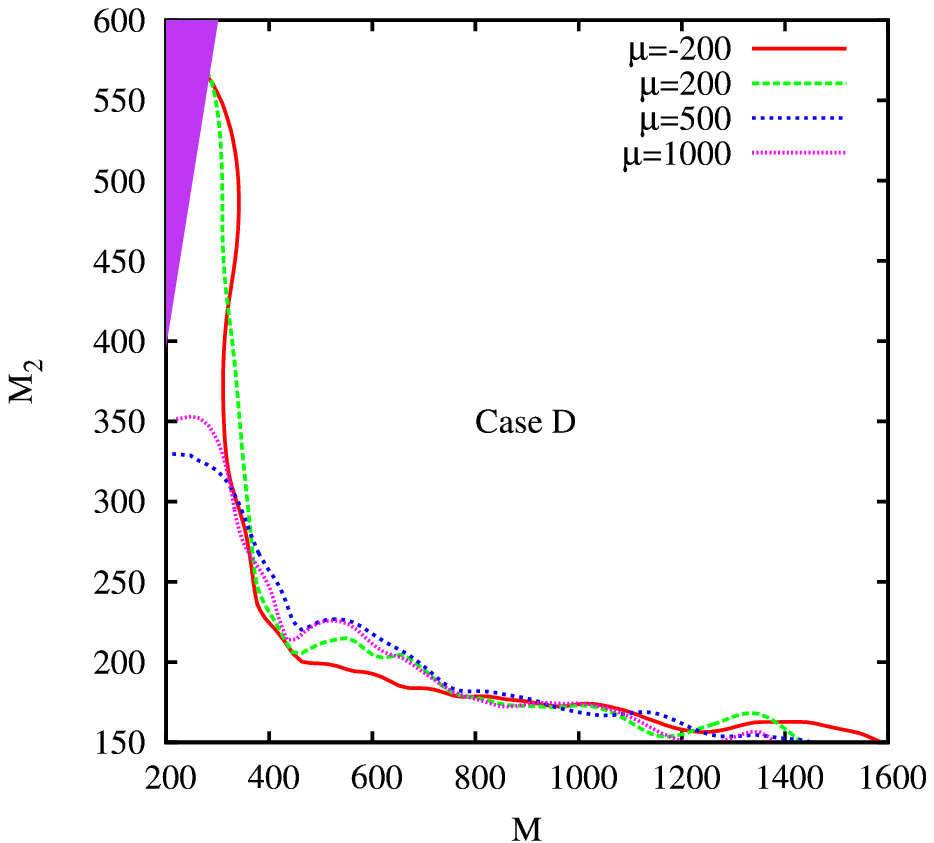} \\ 
\end{tabular}
\caption{\label{fig:mudep} The dependence of the exclusion curves on 
different values of $\mu$ for $\tan \beta=$10.  The least dependence
is for the case where the sbottom is the lighter squark.  The shaded
region at the top in the fourth panel corresponds to stop LSP and is
therefore ruled out. The x-axis refers to $M_{\st_1}$ for Cases A, B
and D and $M_{\sb_1}$ for Case C.}
\end{figure*}

\begin{figure*}[htb]
\includegraphics[width=80mm]{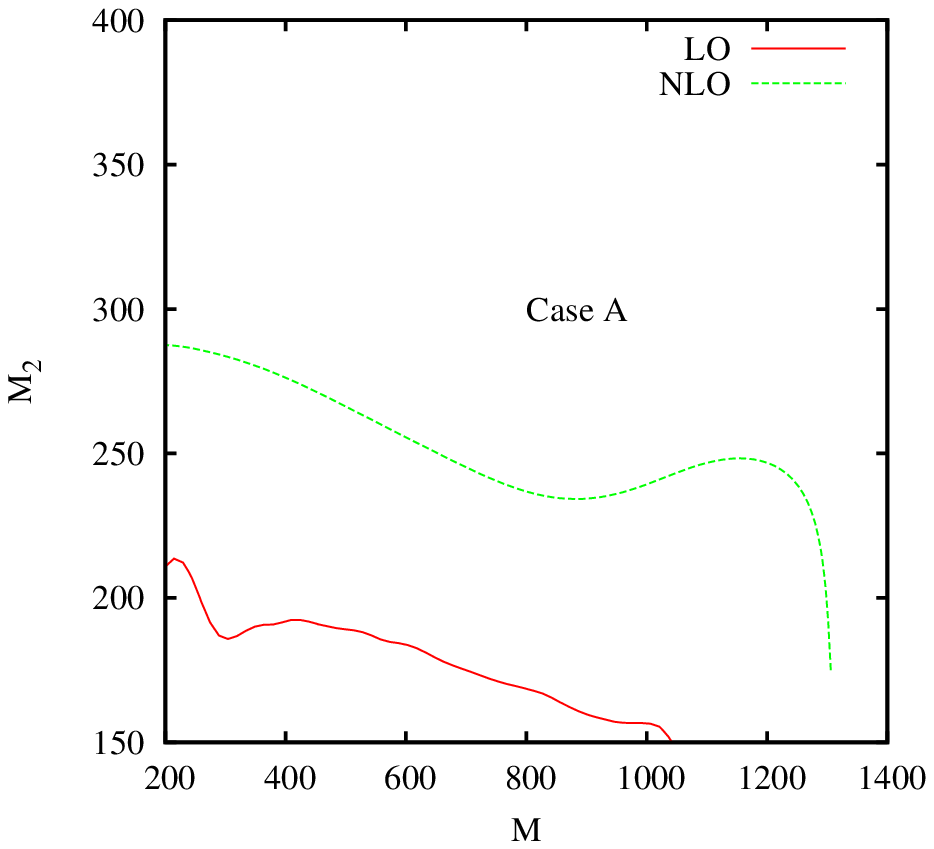}
\includegraphics[width=80mm]{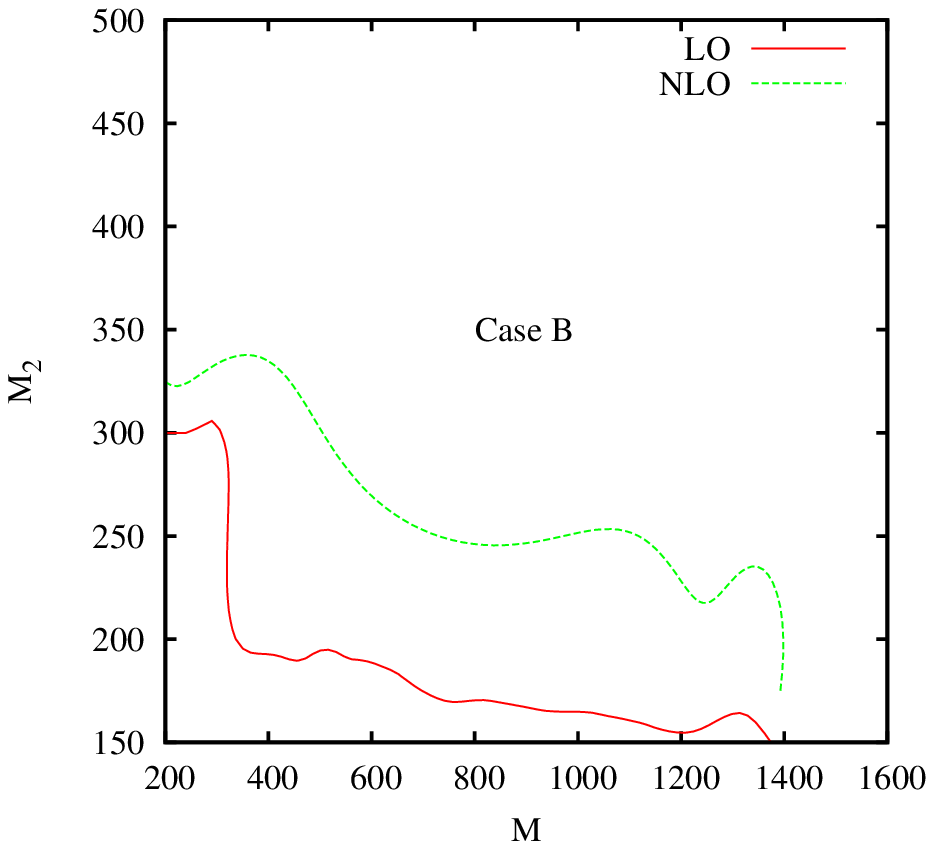}
\includegraphics[width=80mm]{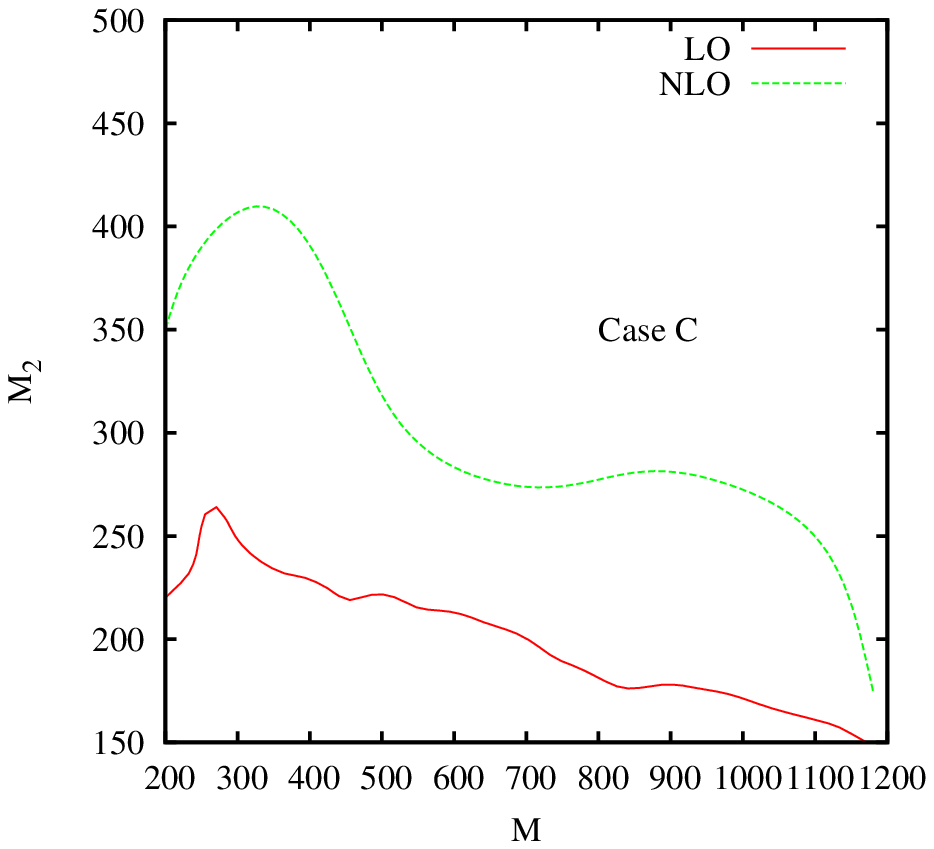}
\includegraphics[width=80mm]{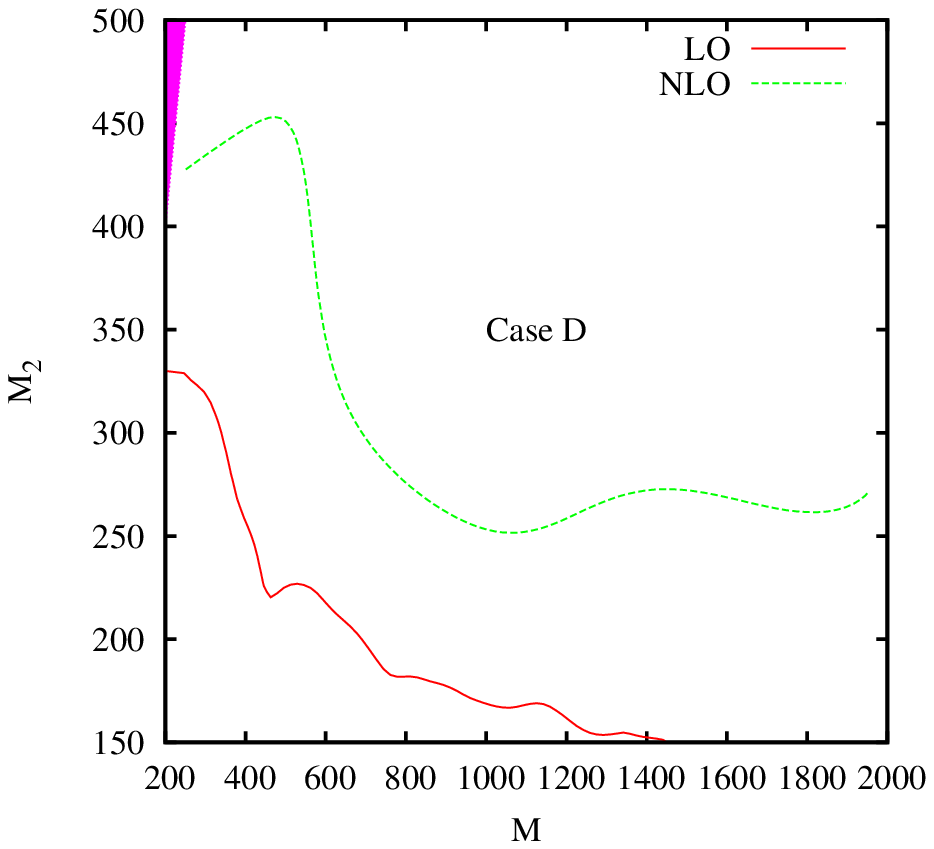}
\caption{\label{fig:st-nlo} Comparison of LO and NLO exclusion curves
  for Cases A, B, C and D, illustrated for values $\tan \beta=10$ and
  $\mu=500$~GeV. The x-axis refers to mass of the lightest third
  generation quark in each case, as discussed in the text.}
\end{figure*}

The hierarchy among $M_1$, $M_2$ and $\mu$ determines the composition
of the neutralino and chargino sector and therefore has strong effects
on the limits.  In particular, for cases with low values of $\mu$, the
lightest neutralino has a significant higgsino fraction.  When $M_2$
becomes large enough, the states $\neut_1$ and $\charg_1$ become
higgsino dominated and their masses remain close to the value of
$\mu$.  In this case, the masses and compositions of $\neut_1$,
$\charg_1$ $\neut_2$ are not affected by changes in the value of
$M_2$, so long it is considerably larger than $\mu$. For large $\mu$,
on the other hand, the allowed decays of the squark will depend
strongly on $M_2$ up to very large values.  Therefore, we expect that
for low squark masses and $\mu$ low with respect to $M_2$, the
exclusion contour is relatively insensitive to $M_2$.  This can
clearly be seen in figure~\ref{fig:mudep} and best illustrated in the
fourth panel corresponding to Case-D.  Here, the production cross
section is high because all four third generation squarks are
degenerate.  Moreover, it can clearly be seen that $\mu = $ -200 and
200 both lead to large exclusion in $M_2$ for small masses.  The
effect is similar also in the panel corresponding to Case-B.  The
third panel, corresponding to sbottom being the lightest shows minimal
change with changing $\mu$.  This is mostly because the decay $\sb_1
\rightarrow b \neut_1$ is always open irrespective of $\mu$ due to the
small mass of the b-quark.  Thus, the sensitivity to $\mu$ is reduced.

The channel with two b-tagged jets and high missing energy (called
$\mathrm{3JD}$ in \cite{ATLAS-CONF-2011-098}) results in most
stringent limits in all cases.  The cross section ruled out in this
case (0.017 pb) is similar to the the number for
$\mathrm{2~jets~+~MET}$ (0.024 pb), however, the latter does not
contribute significantly to the parameter space ruled out.  The reason
for this is two-fold: first, the effective mass cut of 1 TeV for the
jets+MET final states has a much smaller efficiency for small stop
masses.  Secondly, the missing energy cut for $\mathrm{3JD}$ is a
smaller, fixed number ($130~\mathrm{GeV}$) as compared to the
$M_\mathrm{eff}$ dependent one ($\mathrm{MET}> 0.3 M_\mathrm{eff}$) in
the $\mathrm{2~jets~+~MET}$ case.  The reduction of effeciency when
demanding 2 b-tagged jets is not very severe in $\glue \glue$
production events due to four b-quarks in the final state.  Even in
the regions where exclusion is dominated by $\sq \sq^*$-type
production, we still find that the $\mathrm{3JD}$ channel results in
the best exclusions.

To estimate the effect of enhancement due to NLO corrections in the
production cross section, we present the comparison of LO and NLO
curves for each case with $\tan \beta=10$ and $\mu=500$~GeV in
Fig.\ \ref{fig:st-nlo}.  The large k-factors ($\sim 2.5$) in most of
the parameter space result in much stronger limits from the NLO
curves.  However, taking note of the results of our cMSSM limits in
Section~\ref{sec:simulation}, where the LO limits are closer to actual
ATLAS limits, we take the conservative approach of presenting LO
limits for our study.

\subsection*{Results: Case A}

\begin{figure*}
\includegraphics[width=80mm]{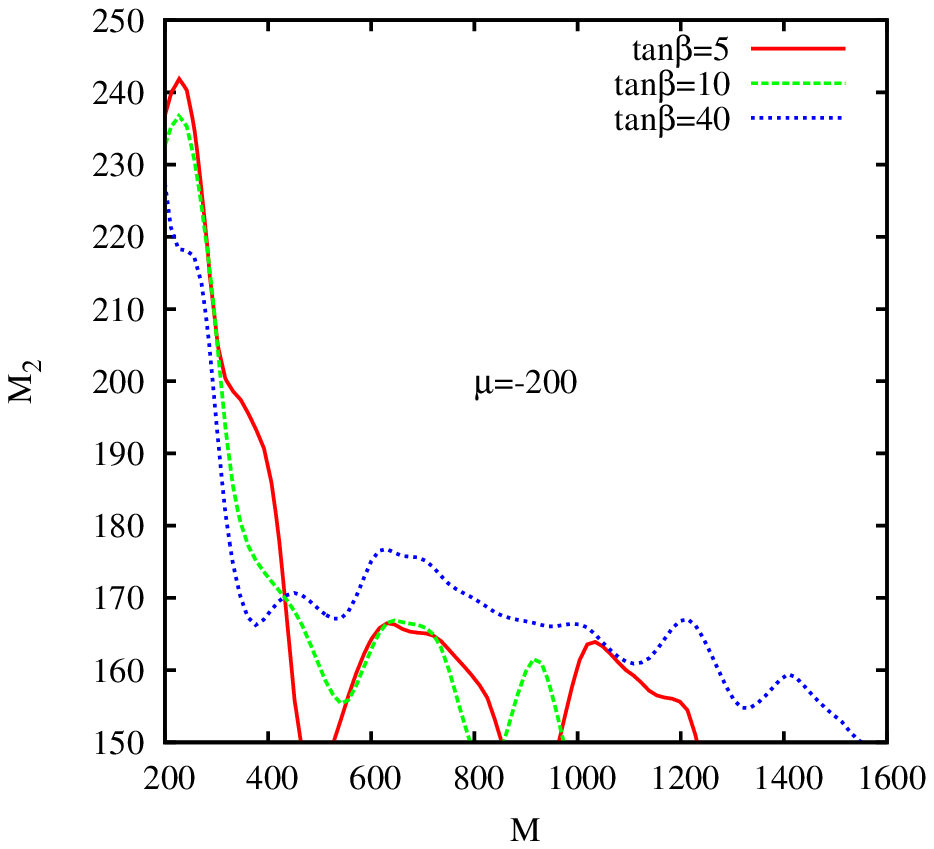}
\includegraphics[width=80mm]{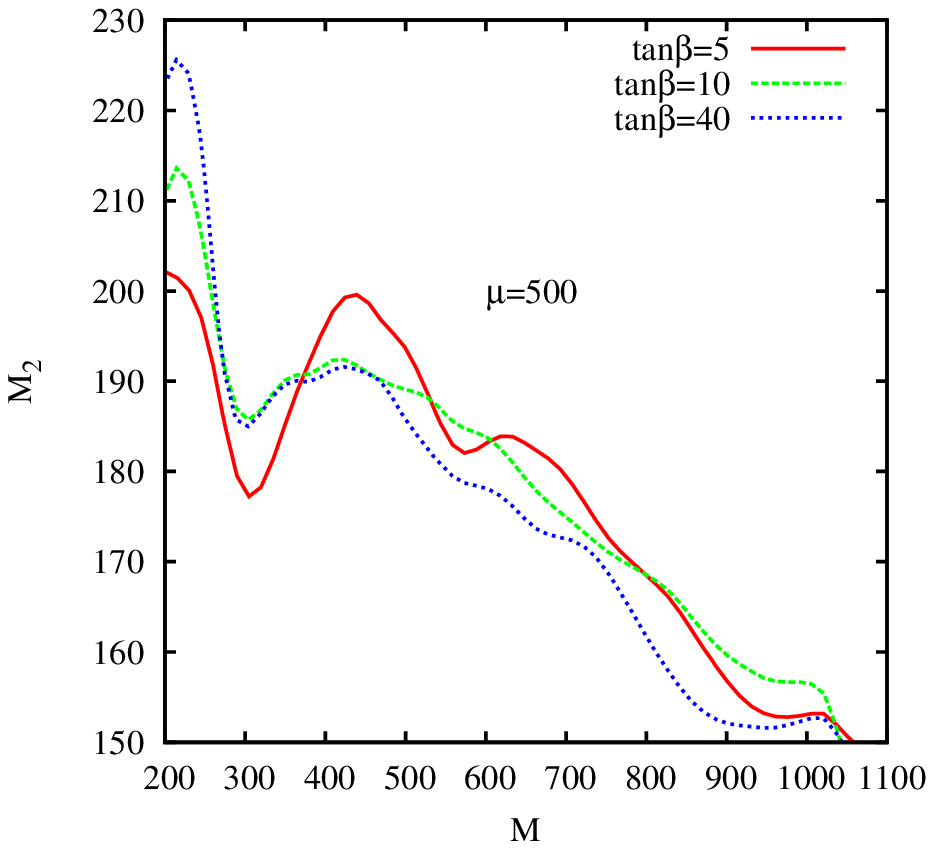}
\caption{\label{fig:st-tanb} Exclusion curves for Case-A (stop lightest scenario).  The x-axis refers to the mass of $\st_1$ squark.}
\end{figure*}

This is the case closest to mSUGRA-type models where $\st_1$ is the
lightest squark. The primary production processes in this case are
$\glue \glue$ and $\st_1 \st_1^*$.  Since our scan starts with $M_2 =$
150 GeV, i.e. a gluino mass of 450 GeV, the decay $\glue \rightarrow t
\bar t \neut_1$ is always open and forms the dominant decay mode.  In
the low mass regions which are probed by the LHC data, the dominant
decay mode of the $\st_1$ depends on the mass hierarchy of $\st_1$,
$\neut_1$ and $\charg_1$.  Since we are working in R-parity conserving
models, we disallow the region where the lighter stop is the lightest
supersymmetric particle (LSP) i.e.\ the region where $M_{\st_1} <
M_{\neut_1}$.  Following this, the hierarchy $M_{\neut_1} < M_{\st_1}
< M_{\charg_1}$ results in the case of stop NLSP (next-to-LSP).  The
dominant mode is $\st_1 \rightarrow t \neut_1$ if $M_{\st_1} > M_t +
M_{\neut_1}$.  Otherwise, we expect $\st_1$ to decay via three-or
four-body decays or via the  mode $\st_1 \rightarrow c \neut_1$.
For this work, we assume that this last mode dominates over the three
or four body decays.  Finally, in the case where $M_{\st_1} > M_b +
M_{\charg_1}$, the decay into $b \charg_1$ is also open.  For larger
stop masses, decays into other neutralinos, second chargino or gluino
may also open.

The effect of various values of $\tan \beta$  for two values of $\mu$
are shown in Fig.\ \ref{fig:st-tanb}.  The exclusion curves show minor
dependence on $\tan \beta$.  We find that for low values of $\mu$, low
$\tan \beta$ results in a larger reach in $M_2$ whereas high $\tan
\beta$ results in a larger reach in $M_{\st}$.  This trend is reversed
for large values of $\mu$, as can be seen in the panel corresponding
to $\mu=500$ GeV of Fig.\ \ref{fig:st-tanb}.  However, it must be
reiterated, that this variation cannot be considered experimentally
significant due to the uncertainties on the exclusion curves in our
analysis.

\subsection*{Results: Case B}
\begin{figure*}
\includegraphics[width=80mm]{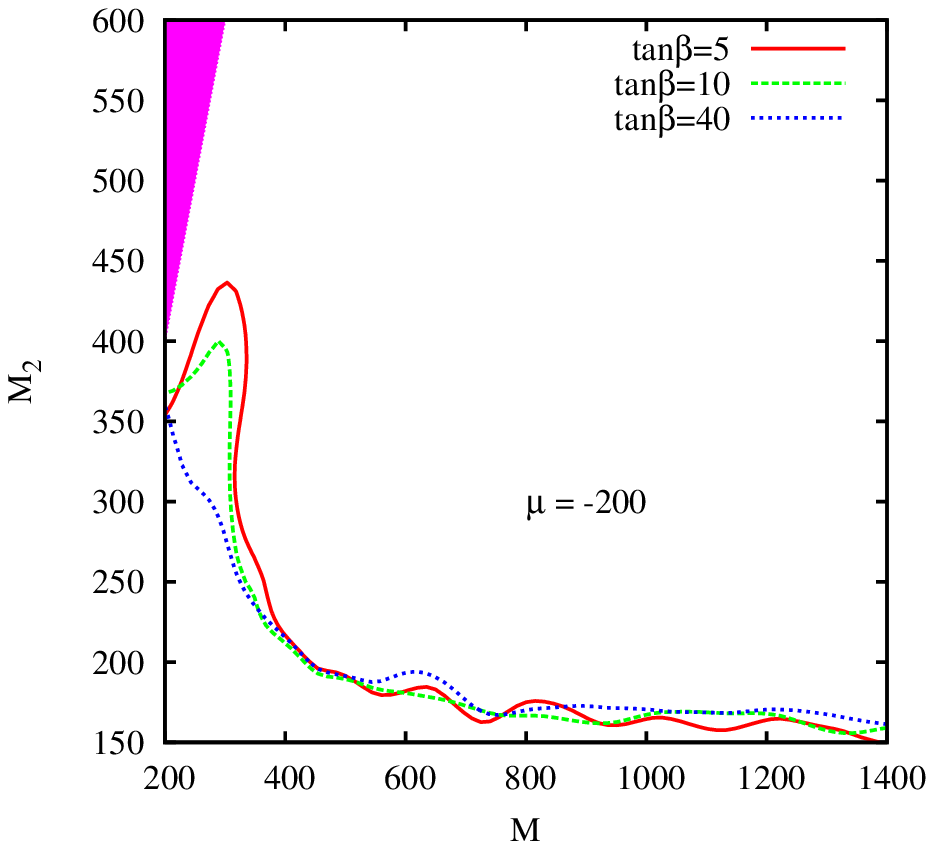}
\includegraphics[width=80mm]{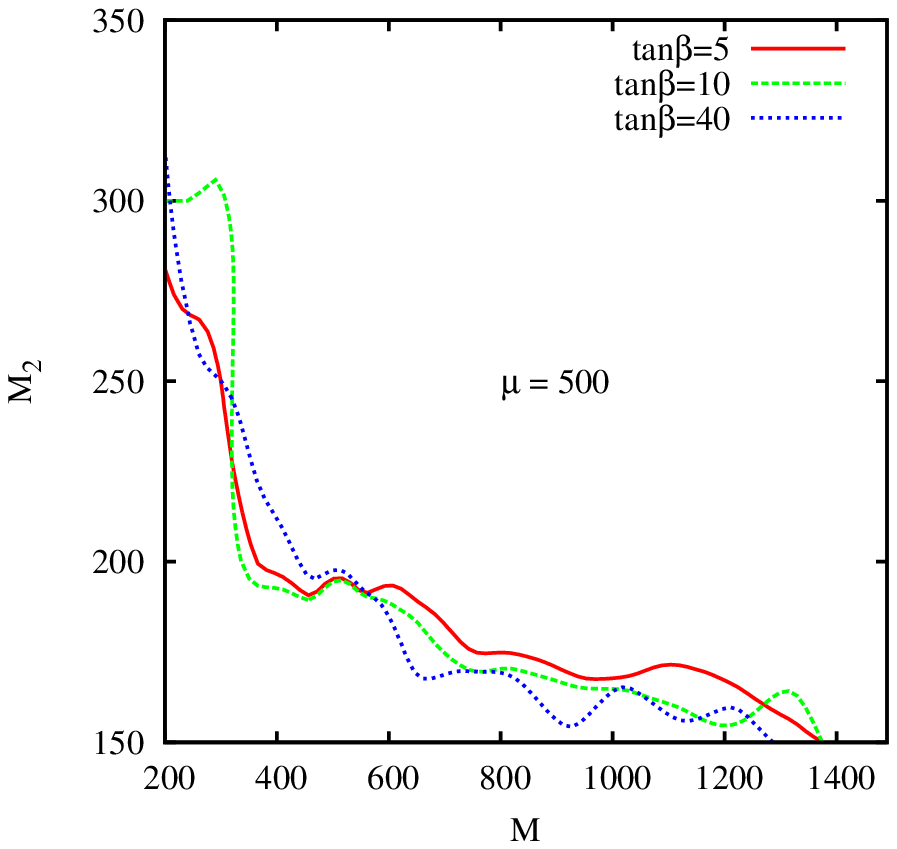}
\caption{\label{fig:lh-tanb} Exclusion curves for Case-B (left-handed degenerate 
stop and sbottom scenario).  The x-axis refers to the common mass of
$\sb_1$ and $\st_1$ squarks.  The shaded region at the top in the
fourth panel corresponds to stop/sbottom LSP and is therefore ruled
out.}
\end{figure*}

In this case, the $\st_1$ and $\sb_1$ form a degenerate pair of
lightest squarks.  They are both primarily left handed and therefore
have an enhanced coupling to wino-like states.  Again, due to the
requirement of neutral dark matter candidate, we disallow any region
with stop or sbottom LSP.  The decays of the lighter stop are similar
to those in case A.  The decay of the lighter sbottom into $b \neut_1$
is almost always allowed and will form the dominant decay for most of
the low mass region.  In cases of large $\mu$, where $M_{\neut_2}
\simeq M_2$, the decay $\sb_1 \rightarrow b \neut_2$ will dominate
over $\sb_1 \rightarrow b \neut_1$ and similarly for stop decays.  The
decay $\sb_1 \rightarrow t \charg_1$ is relatively disfavoured due to
large top mass.  The gluino decays dominantly to $b \bar b \neut_1$ in
the region $M_{\glue} < M_{\sb_1}$ and to $b \sb_1$ otherwise.  The
decays to corresponding top-sector are again disfavoured due to large
top mass.

The dependence of the exclusion curves on $\tan \beta$ is shown in
Fig.\ \ref{fig:lh-tanb}.  For the case of $\mu=-200$ GeV and $\tan
\beta = 5$, we find that $\st_1$ and $\sb_1$ up to 300 GeV are ruled
out for gluino masses up to 700 GeV.  The region just below the
$\st_1$ or $\sb_1$-LSP region is still allowed, as  the
near-degeneracy of their mass and the mass of the LSP results in a low
missing energy and $M_{\mathrm{eff}}$ spectrum which does not satisfy
the hardness cuts imposed.

\subsection*{Results: Case C}

\begin{figure*}
\includegraphics[width=80mm]{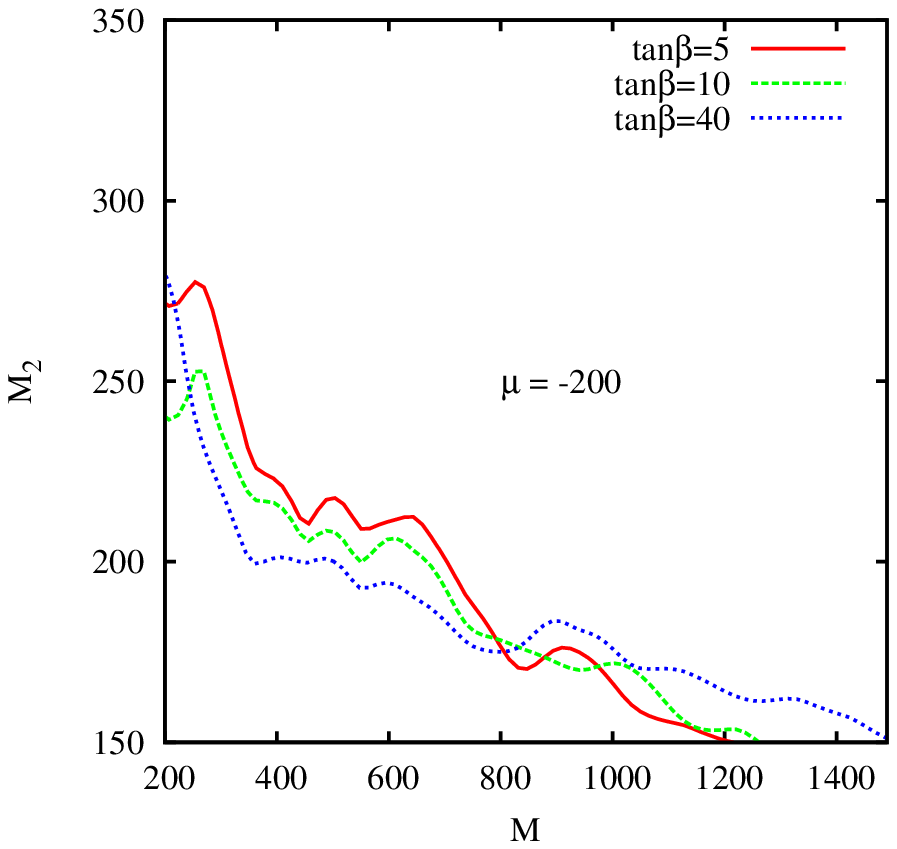}
\includegraphics[width=80mm]{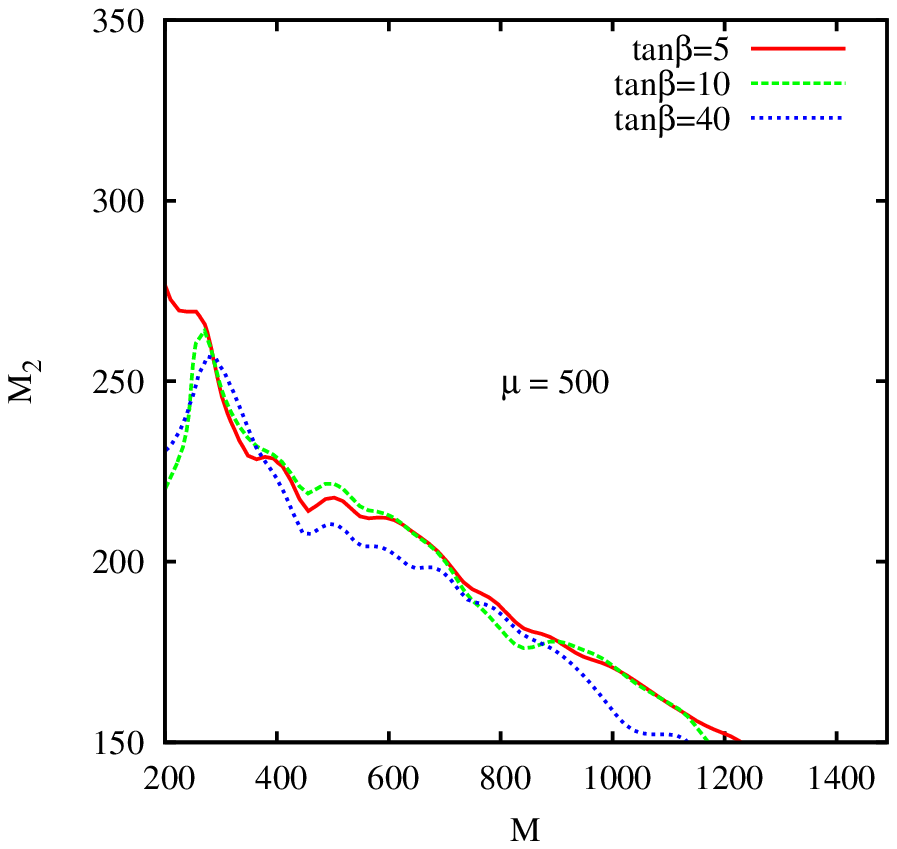}
\caption{\label{fig:sb-tanb} Exclusion curves for Case-C (sbottom lightest scenario).  The x-axis refers to the mass of $\sb_1$ squark.}
\end{figure*}

This case considers the situation where both the stop states are
heavier than the lightest sbottom state.  The primary production
processes are $\glue \glue$ and $\sb_1 \sb_1^*$.  As before, we
disallow the region $M_{\sb_1} < M_{\neut_1}$.  The region
$M_{\neut_1} < M_{\sb_1} < M_{\charg_1}$ corresponds to a $\sb_1$-NLSP
with the dominant decay $\sb_1 \rightarrow b \neut_1$ .  The gluino
dominantly decays via $b \bar b \neut_1$ in the region $M_{\glue} <
M_{\sb_1}$ and to $b \sb_1$ otherwise.  This case is the closest to
the scenarios considered by the ATLAS collaboration for the
interpretation of their b-jet and missing energy searches.  They split
their analysis into a case where they disallow any three body decays
of the gluino via $\glue \rightarrow b \bar b \neut_1$ and the case of
a simplified model where there are no two-body decays but all decays
are via this channel.  According to the first case, they rule out
gluino masses up to 720 GeV for $\sb_1$ masses up to 600 GeV
\cite{ATLAS-CONF-2011-098}.  As can be seen from Fig.\
\ref{fig:sb-tanb}, for $\sb_1 =$ 600 GeV, we rule out $M_{\glue} < $
570-660 GeV for $\mu = -200$ and $M_{\glue} < $ 600-660 GeV for $\mu =
500$ GeV.

\subsection*{Results: Case D}

\begin{figure*}
\includegraphics[width=80mm]{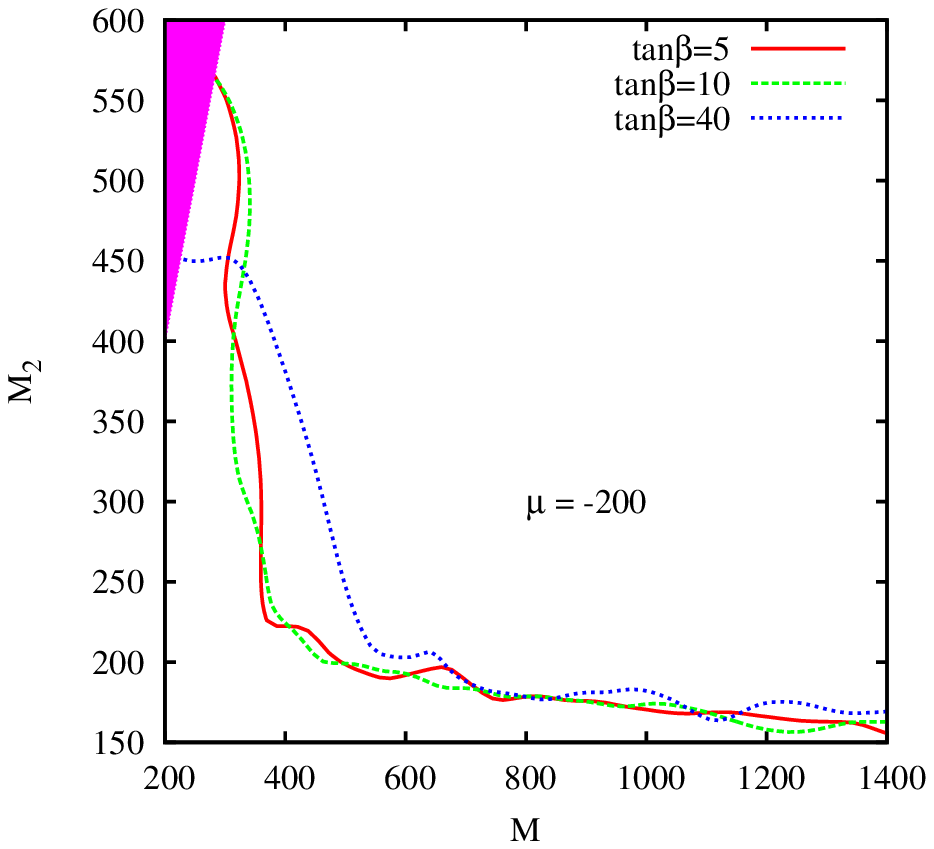}
\includegraphics[width=80mm]{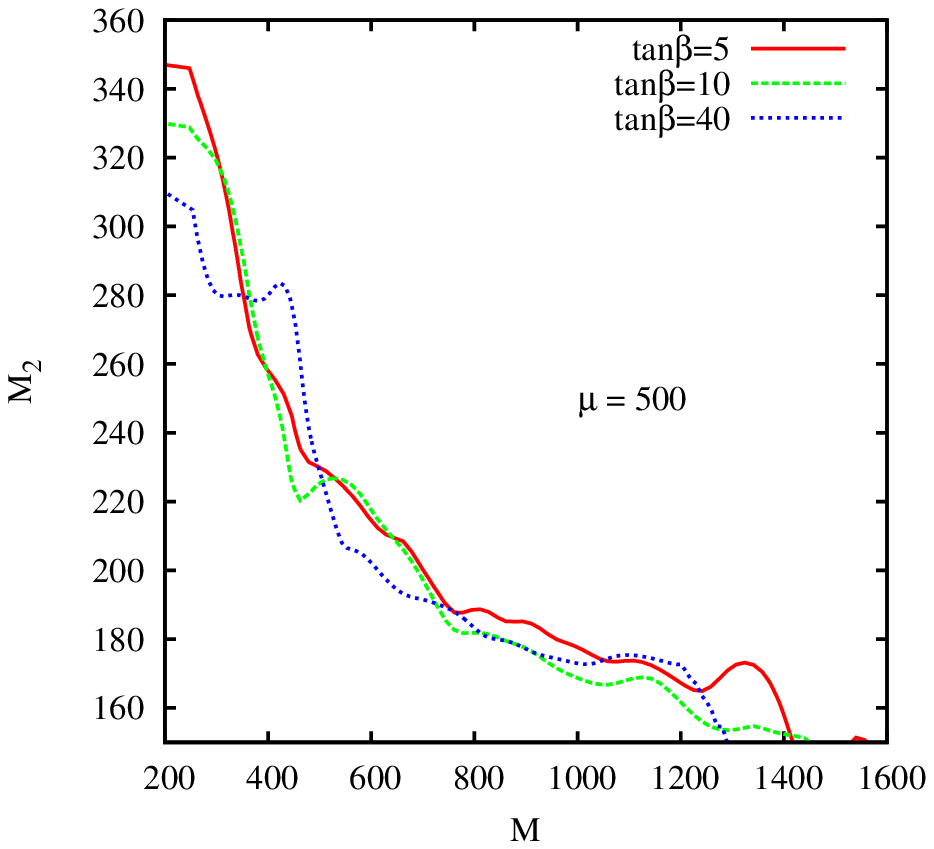}
\caption{\label{fig:eq-tanb} Exclusion curves for 
Case-D (maximal mixing scenario).  The x-axis refers to the common
mass of all the third generation squarks.  The shaded region at the
top corresponds to stop LSP and is therefore ruled out.}
\end{figure*}

In the maximal mixing scenario in both the stop and sbottom sector,
all four squarks of the third generation have nearly degenerate
masses.  Therefore the production cross section is maximum for this
scenario and the limits are strongest.  The decay scheme for the
sbottom is same as in Case B whereas for the stop, it is the same as
in Case A.  The gluino can now decay both via stops or sbottoms, but
the large mass of the top means it decays preferentially via $b\sb_i$
channels.  As expected, low $\mu$ results in large exclusion in $M_2$
for low squark masses.  The dependence of the limits on $\tan \beta$
is shown in Fig.\ \ref{fig:eq-tanb}.  The case for $\mu = -200$ GeV
(and similarly $\mu=200$ GeV) result in an exclusion of third
generation squark masses of 280 GeV for all allowed values of $M_2$.
This case is similar to the one considered in \cite{Papucci:2011wy},
where, for naturalness requirements they require $\st_R, \st_L$ and
$\sb_L$ to be degenerate and assume a Higgsino LSP.  Their limits on
the mass of the third generation squarks lie between 200-300 GeV.

The exclusion limits for the case of $\mu=500$~GeV does not show any
exclusions that are independent of gaugino (and hence gluino) masses.
The third-generation squark masses are completely un-constrained for
$M_{\glue}>1$~TeV.  Whereas, the approximate requirement for
naturalness that $M_{\st_1} <500$~GeV would translate into $M_{\glue}
>600$~GeV.

In conclusion, among the four cases discussed above, only the case
with degenerate third-generation squarks and low-$\mu$ leads to mass
limits independent of gluino mass -- that of 280 GeV.  In most other
cases, we find that limits depend strongly on the composition of the
neutralinos and charginos.  The case where only $\st_1$ is accessible
is the least constraining, mainly due to low production cross sections
compared to the other cases.  For the case where the LSP is a almost
pure Bino (high-$\mu$), $M_{\st_1} = 200$ GeV is ruled out for a
gluino mass less than 570 GeV.  Taking into account all values of
$\mu$ and $\tan \beta$, $M_{\st_1} = 200$~GeV is ruled out for gluino
masses in the range $570-720$~GeV.

The case of lightest third generation squark being $\sb_1$ is the most
insensitive to variations of both $\mu$ and $\tan \beta$.  For this
case, our limits are consistent with ATLAS's own interpretations
within 10\%.  $M_{\sb_1}=200$~GeV is ruled out for gluino masses
between $680-820$~GeV.  The case of degenerate left-handed squarks
rules out $M_{\st_1}=M_{\sb_1}=200$~GeV for gluino masses in the range
$900-1050$~GeV.  And finally, the case with all squarks degenerate
rules out $M_{\st,\sb}=200$ for $M_{\glue}<900$~GeV in the worst case,
and for all gluino masses in the best case.

\section{High-scale non-universal scalar scenarios}
\label{sec:high-scale}

Besides the low-scale study done in the previous section, it is also
possible to perform a scan over high-scale parameters.  The advantage
of a high-scale analysis is that the hierarchy of the particles is
uniquely and consistently determined from the renormalisation group
(RG) running of masses to low scale from given high-scale parameters.
We use the simplification afforded by this model to include the
effects of sleptonic sector in our analysis.

It is possible that the limits described in the previous section are
diluted if slepton masses are allowed to be light.  This is because
the gauginos would then decay predominantly to sleptons resulting in
leptonic final states which would be discarded since all the analyses
considered here have a lepton veto.  Including the full lepton sector
in the low-scale pMSSM requires five more parameters and makes a
general study far more complicated.  We therefore leave a fully model
independent investigation of interpreting the ATLAS limits involving a
low mass slepton sector to a future work.  However, we partially answer the
question as to whether the limits are diluted by studying some illustrative
cases, as described below.

Even though the soft-scalar masses may in principle take separate
values, the constrains from flavour changing neutral currents (FCNC)
from meson decays dictate that the first two generation squarks remain
degenerate.  Similarly, absence of decays like $\mu \rightarrow \gamma
e$ means that the masses of first two generations of sleptons also
have to degenerate.  Therefore we can consider three schemes of
non-universality: 
\begin{itemize}
\item \textbf{Case  HA: } Third generation squarks are lighter than all
other sfermions.  This will lead to a hierarchy similar to Case A in the preceding analysis.

\item \textbf{Case HB:} Third generation squarks and sleptons are
  lighter than all other sfermions.  This leads to light staus and
  tau-sneutrinos.  Possibly, this would lead to $b\tau$ final states
  which have been studied in \cite{Bhattacharyya:2011se}.

\item \textbf{Case HC: } Third generation squarks and all sleptons are
  light.  Comparing the limits in this case to those in Case HA will
  answer the question of dilution of limits due to leptonic
  signatures.
\end{itemize}

\begin{figure*}[ht]
\includegraphics[width=80mm]{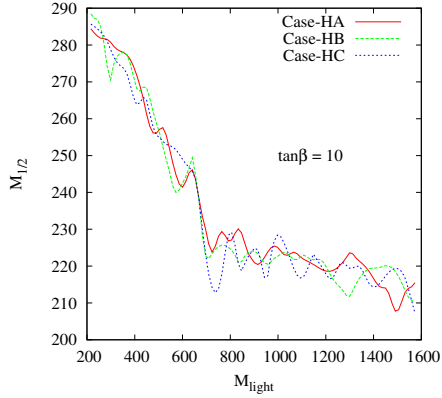}
\caption{\label{fig:high-scale-10} Comparison of exclusion curves from 
three high-scale non-universal scenarios.  All cases agree with each
other within statistical uncertainties.}
\end{figure*}

As in the previous analysis, we retain the cMSSM-like gaugino masses.
We now have two mass scales in the scalar sector -- the scale of
heavy, decoupled particles ($M_{\mathrm{heavy}}$ = 2 TeV) and the
scale of light scalars ($M_{\mathrm{light}}$).  We once again assume
that the gaugino sector follows the universal structure and we set the
higgs mass parameters for the two higgs doublets to be same as
$M_{\mathrm{light}}$.  The parameters are all set at the grand unified
theory (GUT) scale and the TeV-scale values are determined by RG
running using the program SUSPect.  The exclusion curves obtained for
the three cases are shown in Fig.\ \ref{fig:high-scale-10} for a value
of $\tan \beta=$ 10.  The results for values of $\tan \beta=$ 5, 40
are similar.  We do not see any significant difference among the
exclusion curves for the various high-scale cases.  This can be
interpreted as the robustness of the zero-lepton signals against
different slepton masses and justifies the assumption of decoupled
slepton masses made in the previous section.

\section{Conclusions}
\label{sec:conclusions}
We have investigated the consequences of the recent ATLAS data in
channels with (b-)jets and missing energy on the limits on the mass of
the third generation squarks.  We work in the pMSSM framework, with
TeV-scale parameters, without requiring a high-scale breaking scheme.
For obtaining relatively model independent limits on the third
generation squark masses, we decouple the first two squark generations
as well as all sleptons.  We also explicitly show that decoupling of
sleptons is not likely to affect the limits as long as we work with
0-lepton signatures.  We find that a stop of mass 200 GeV can be ruled
out for a gluino mass of 570 GeV in the least constraining case
whereas a stop of mass 500 GeV is allowed for gluino masses upward of
450-880 GeV depending on the structure of the third-generation squark
sector, and the parameters $\mu$ and $\tan \beta$.  In the case where
all third generation squarks are degenerate, we can rule out masses
less than 280 GeV for $|\mu| \leq 200$~GeV, independent of the gluino
mass.

One may also question the assumption of the gaugino mass pattern of
$M_1:M_2:M_3 \simeq 1:2:6$.  We can expect the case of right-handed sbottom
being the lightest of the third generation to be fairly independent of
the assumption of gaugino mass pattern since the only dominant decays
are $\sb_1 \rightarrow b \neut_1$ and $\glue \rightarrow b \bar b
\neut_1$, and both are always allowed (except for very compressed
spectra).  We would expect significant deviations from the stop limits
when, for example, the decay $\st_1 \rightarrow b \charg_1$ is largely
inaccessible because both $M_2$ and $\mu$ are so high that the
charginos are generally heavier than the stop.  In this case, the
dominant decay for most of the parameter space would be $\st_1
\rightarrow t \neut_1$.  In the current study, regions where the $t
\neut_1$ decay was kinematically disallowed was still largely covered
by the $b \charg_1$ decay, thus leaving only a small region of
parameter space corresponding to the flavour violating decay.
However, in the absence of the decay into a chargino, one would need
to examine in detail, the relative strengths of the (highly
model-dependent) flavour-violating decay $\st_1 \rightarrow c \neut_1$
and the three-body decay of the stop.  In other cases, where the stop
still decays via standard channels, we do not expect significant
deviations from our limits.

\noindent \textbf{Acknowledgements:}  We thank Bruce Mellado for information 
on smearing parameters for detector effects.  This work was partially
supported by funding available from the Department of Atomic Energy,
Government of India for the Regional Centre for Accelerator-based
Particle Physics, Harish-Chandra Research Institute. Computational
work for this study was partially carried out at the cluster computing
facility of Harish-Chandra Research Institute ({\tt
http:/$\!$/cluster.mri.ernet.in}).

\bibliography{paper}
\end{document}